\documentclass[a4paper,12pt]{article}
\usepackage{jheppub} 
\usepackage{subfigure}
\usepackage{graphicx}
\usepackage{color}
 
\usepackage{soul}
\usepackage{amsmath,amsfonts,amssymb,amsthm}
\usepackage{tikz}
\usepackage{pifont}
\usepackage{bbding}
\usepackage{multirow}
\usepackage{tabularx, makecell, booktabs}

\usepackage{hyperref}

\makeatletter
\gdef\@fpheader{Accepted in JHEP\\
\footnotesize{DOI: 10.1007/JHEP02(2025)043}}
\makeatother

\title{Chaos Bound and its violation in Black p-brane}
\author{Pinaki Dutta,}
\author{Kamal L. Panigrahi,}
\author{Balbeer Singh}
\affiliation{Department of Physics, Indian Institute of Technology Kharagpur,\\ Kharagpur 721 302, India}

\emailAdd{coolguddu0815@kgpian.iitkgp.ac.in}
\emailAdd{panigrahi@phy.iitkgp.ac.in}
\emailAdd{curiosity1729@kgpian.iitkgp.ac.in}

\abstract{In this work, we have extensively investigated the dynamics of circular geodesic (chargeless massive particle) followed by the investigation of the pulsating classical string in the p-brane background. This study is a continuation of our previous work \href{https://link.springer.com/article/10.1007/JHEP10(2023)189}{JHEP10(2023)189}, 
in which we numerically identified the presence of chaos for a classical string hovering near generic p-branes ($p < 7$). Here, for a particle probe, we have found evidence of chaos in the vicinity of the horizon. Furthermore, we observed a violation of the well-known MSS bound in specific extremal p-branes; however, no such violation is seen in the non-extremal cases. Similar observations were made for the classical string, where the violation of the bound is significant near the horizon. Thus, our semi-analytical arguments demonstrate that chaotic dynamics in black p-branes exhibit the (generalized) universal bound with notable violations, regardless of whether a particle or classical string is used as a probe.}  

\keywords{Black holes in String Theory, Bosonic Strings, Integrable Field Theories, P-Branes.}

\begin{document}
\maketitle
\flushbottom
\section{Introduction}
Integrability plays a pivotal role in theoretical physics. Various branches of physics share the feature of integrability. Even though integrable models are rare to find, their importance lies in the underlying symmetries of such models, and therefore become exactly solvable. More specifically, Liouville integrability suggests that the motion of the dynamical system is exactly solvable if all the submanifolds of the phase space are integrable \cite{Beisert:2010jr,Torrielli:2016ufi, Zarembo:2017muf}. \\
In recent years, the framework of integrability has been effectively utilized within the context of AdS/CFT correspondence \cite{Maldacena:1997re}. A lot of progress has been made in this direction in the last two decades, see \cite{Babichenko:2009dk, Bombardelli:2016rwb,van2014integrability,klimcik2008integrability,delduc2014integrable,matsumoto2014lunin} and the references therein. One of the important questions concerns the criterion for integrability in the gauge field theories. To this end, one of the celebrated approaches in this direction is to use the techniques of analytic non-integrability, as first done in \cite{PhysRevD.84.046006,Basu:2011di}. Instead of claiming the infinite conserved charges and the Lax pair connection thereof which is rather a cumbersome task to achieve \cite{PhysRevD.69.046002}, the idea of analytic non-integrability lies in the fact to disprove the integrability of the $\sigma$-model in the suitable choice of the ansatz for the string configuration of the given system. This can be done by introducing small variations to the corresponding equation of motion known as normal variational equations (NVE)\cite{ruiz1999differential,yagasakiGaloisianObstructionsIntegrability2003}, which cannot be solved exactly by the method of quadratures. To this end, the Kovacic algorithm helps to demonstrate the non-existence of the Liouville solutions \cite{kovacicAlgorithmSolvingSecond1986}.    \\
\\
The relation between chaos and non-integrability is yet another field which has gained much attention in recent times \cite{PhysRevD.102.106022,Wulff:2017vhv,Wulff:2017hzy,Nunez:2018ags,McLoughlin:2022jyt,Pal:2023ckk,Pal:2023kwc,Pal:2023bjz,Dutta:2023yhx} and manifest the fact that chaos implies non-integrability. The studies of the chaotic motion and non-integrability suggest expanding the dictionary of the AdS/CFT correspondence \cite{PandoZayas:2010xpn}, however, the picture of the present status of the art seems to be slightly unclear. Yet another direction worth exploration is to look at the relation and the source of the chaos/non-integrability in view of the celebrated Maldacena Shenker Stanford(MSS) bound \cite{Maldacena:2015waa}:
\begin{equation}
    \lambda \le 2 \pi T
\end{equation}
where $T$ denotes the Hawking temperature and $\lambda$ is the Lyapunov exponent. The MSS bound was originally proposed by considering shock waves near the horizon and AdS/CFT correspondence. However, it has been argued that the bound is universally present in the plethora of (semi-) classical as well as quantum systems such as black holes which are considered to be the fastest scramblers in nature\cite{dalui2019presence,swingle2016measuring,shenker2015stringy,shenker2014black,shenker2014multiple,susskind2018things,polchinski2015chaos}. In light of the point particle dynamics, various studies of black holes indicate the existence of chaos \cite{suzuki1997chaos,sota1996chaos,bombelli1992chaos,li2019chaotic,hartl2003dynamics}. In recent years, the presence of chaotic motion has been attributed to the unstable equilibrium circular orbits in the near-horizon geometry of black holes \cite{kan2022bound,gwak2022violation,zhao2018static,gao_chaos_2022,xie2023circular,jeong2023homoclinic,wang2023spatial,lei2021chaos,yu2023violating}. Such chaotic motions have been observed for generic perturbations as well. Therefore eventually in \cite{Hashimoto:2016dfz}, it was proposed that there exists a universal bound for the Lyapunov exponent $\lambda$ of chaotic motions induced by black holes given by
\begin{equation}
    \lambda \le \kappa
\end{equation}
where $\kappa$ represents the surface gravity of the black hole. In many dual-field theories, the bound saturates such as in the SYK model \cite{marcus_new_2019,kitaev2015simple,kitaev2015simple2}, but several works point out the 
 violation of the bound \cite{zhao2018static,gao_chaos_2022, xie2023circular,gwak_violation_2022,jeong2023homoclinic,wang2023spatial,lei2021chaos,yu2023violating,lei2022circular,prihadi_chaos_2023,das2024near,lei2024thermodynamic}. Recently, for closed string dynamics around the AdS black hole, the bound was generalized with the correction of the winding number $n$ \cite{cubrovic_bound_2019}
\begin{eqnarray}\label{eqn: genbound}
     \lambda \le 2 \pi n T
\end{eqnarray}

To our current understanding, there are not many studies \cite{guo2022probing} on the interplay of the geodesic motion and the classical string as a probe in the context of the chaotic motion and the universal bound. In this work, we attempt to shed some light on the subtle interplay of the two probes leading to chaos in the background geometry of the black p-brane, $p<7$.
Our earlier work \cite{Dutta:2023yhx} shows that the closed winding string exhibits the chaotic motion in the black 5-brane and 6-brane geometry. Following a similar trail, here we again consistently truncate the two-dimensional string equations of motion for the pulsating string into the one-dimensional system and establish the non-integrability followed by the presence of chaos, for the suitable choice of the ansatz in such a way that one also extends the dynamics to include the lower-dimensional branes. Our study shows that in the case of circular geodesics, violation of the bound inequality depends primarily on the charge of the p-brane, whereas for the pulsating string, it depends on the interplay of charge and winding number of the string.
It is to be mentioned that our analysis follows in general for all branes: $1 \le p \le 6$, however, to avoid redundancy in the numerical works, for the sake of brevity, we present only certain branes while performing the numerical computation. 

Finally, for completeness, we would like to 
highlight 
on recent notable developments in the study of black hole chaoticity
studied via string scattering amplitudes 
at both classical and quantum levels. Classical scattering of closed string configurations, analyzed using ring strings as probes, identifies fractal unstable orbits to characterize chaotic regimes (\cite{frolov_chaotic_1999,basu_chaotic_2017,Dutta:2023yhx} and see references mentioned within).
In quantum systems, chaotic Hamiltonians are explored through random matrix theory, where the eigenvalue spacings follow Wigner-Dyson statistics. Recent work on highly excited strings (HES) links chaos to erratic scattering amplitudes and eigenphase spacings, motivated by black hole/string correspondence, suggesting an equivalence between highly excited strings and semiclassical black holes at weak coupling \cite{gross2021chaotic, rosenhaus2021chaos, firrotta2022photon, rosenhaus2022chaos,susskind1998some,horowitz1997correspondence,horowitz1998self}.

The study of chaos, scrambling, and universal bounds can potentially be applied to string-theoretic, horizon-scale microstructures known as fuzzballs, which replace black hole solutions \cite{mathur2005fuzzball, mayerson2020fuzzballs}. Furthermore, the framework of random matrices and the Out-of-Time-Order Correlator (OTOC) provides a foundation for investigating quantum chaos in fuzzball geometries \cite{das2023fuzzballs, chen2024bps}. In \cite{bianchi2020chaos}, massless geodesics are used to investigate chaotic scattering near the photon sphere of fuzzball geometries, validating the MSS bound, with the notable bound violation when the photon-sphere is close to the horizon in the extremal black hole scenario. It would be intriguing to study chaotic scattering in specific fuzzball configurations of the Dp-Dq brane system. However, we leave this exploration for future work.
\newline

The structure of the paper is as follows: In section \ref{review}, we briefly review the non-extremal black p-brane. In the next section \ref{particle probe}, we investigate the dynamics of the circular motion of a particle in the vicinity of the p-brane and show the existence of the Lyapunov exponent, followed by a numerical study on the universal chaos bound. 
In section \ref{non-integrability}, we bring the closed circular string in the presence of the black p-brane and first prove the non-integrability by showing that the corresponding NVE possesses no Liouville solution. 
 In section \ref{numerical bound}, we investigate the generalised chaos bound in the closed string dynamics and numerically check the validation of the bound. Finally, in the last section \ref{conclusions}, we conclude our results with some interesting future directions.

\clearpage
\section{Brief review: Non-extremal Black p-brane} \label{review}
Non-extremal black p-branes are the solutions of 10-dimensional low energy string theory \cite{Horowitz:1991cd,becker2006string} and the corresponding metric ($p <7$) is:
\begin{equation}\label{eqn: brane_metric2}
ds^{2} = -\Delta_{+}\Delta_{-}^{-1/2}dt^{2}+ \Delta_{-}^{1/2}\sum_{i=1}^{p}dx_{i}^{2} + \Delta_{+}^{-1}\Delta_{-}^{\gamma}dr^{2}+r^{2} \Delta_{-}^{\gamma +1} d\Omega_{8-p}^{2}
\end{equation}
where $\Delta_{\pm} = 1-(\frac{r_{\pm}}{r})^{7-p}$, $\gamma = -\frac{1}{2} - \frac{5-p}{7-p}$ with
$r_{+}$ and $r_{-}$ representing outer horizon and inner horizon radii respectively. The charge and mass per unit p-volume of the black brane are respectively given by 
\vspace{3mm}
\begin{flalign*}
Q & = \frac{7-p}{2}(r_{+} r_{-})^{(7-p)/2}\\
M & = \frac{\Omega_{8-p}}{2k_{10}^2} \Big( (8-p)r_{+}^{7-p} - r_{-}^{7-p}  \Big)
\end{flalign*}
where $\Omega_{8-p}$ is the volume of unit (${8-p}$)- sphere and $k_{10}^2$ = $8\pi G_{10}$. The metric (\ref{eqn: brane_metric2}) can be re-written as 
\begin{equation}\label{eqn: abel}
 ds^{2} = H^{-1/2} \Big( -fdt^{2}+\sum_{i=1}^{p} dx_{i}^{2} \Big) + H^{1/2} \Big( f^{-1}dr^{2}+r^{2}d\Omega_{8-p} \Big)
 \end{equation}
via the transformations:
\begin{center}
 $r^{7-p} = \Tilde{r}^{7-p} + r_{-}^{7-p}$, \hspace{3mm}$\Tilde{r}_{+}^{7-p} = \mu ^{7-p} \cosh^{2}\beta $, \hspace{3mm} $\Tilde{r}_{-}^{7-p} = \mu ^{7-p} \sinh^{2}\beta$ 
 \end{center} 
 where $H = 1 + (\frac{r_{-}}{r})^{7-p}$ and $f = 1-(\frac{\mu}{r})^{7-p}$. Writing  $ \mathcal{F}$ = $H^{-1/2} f$, the equation \ref{eqn: abel} becomes:
 \begin{equation}\label{eqn: cuvrobic}
     ds^{2} = - \mathcal{F}(r) dt^{2} + \frac{1}{\mathcal{F}(r)} dr^{2} + H^{1/2} r^{2}d\Omega_{8-p} + H^{-1/2} \sum_{i=1}^{p} dx_{i}^{2} 
 \end{equation}
The extremal black brane solutions can be obtained by setting $r_{+} = r_{-}$ in the equation \ref{eqn: brane_metric2}. 
Also, with the analytic continuation of time co-ordinate, we obtain the inverse temperature \cite{Ohshima:2005ha, duff_black_1996}
 \begin{flalign*}
     \frac{1}{T} & = \frac{4\pi\mu \cosh\beta}{7-p}\\
    or \hspace{3mm}  \frac{1}{T}&  = 2 \pi \Big( \frac{2 r_{+} }{7-p} \Big[ 1- (\frac{r_{-}}{r_{+}})^{7-p} \Big ] ^{\frac{-5+p}{2(7-p)}}     \Big)
 \end{flalign*}
and therefore the expression for surface gravity is given by
\begin{align} \label{kappa}
   \kappa = \frac{(7-p)}{2 r_{+}} \left(1-\left(\frac{r_{-}}{r_{+}}\right)^{7-p}\right)^{-\frac{p-5}{2(7-p)}}
\end{align}
 
 \section{Probe particle around the black p-brane}\label{particle probe}
In the literature, various techniques exist to determine the Lyapunov exponent, such as those presented in \cite{pradhan2016stability, cardoso2009geodesic,gao_chaos_2022}. Using the Lagrangian framework, Lyapunov exponents can be determined for particle motion. The particle's equations of motion are written as:
\begin{align}
    \frac{d y^{i}}{d t} = F_{i}(x^{j}).
\end{align}
When linearizing these equations around a particular orbit:
\begin{align}
    \frac{d \delta y^{i} (t)}{dt}= K_{ij} \delta y^{j}(t),
\end{align}
where $K_{ij}$ is the Jacobian matrix defined as 
\begin{align}
    K_{ij}= \frac{\partial F_{i}}{\partial y^{j}},
\end{align}
the solution can be expressed as 
\begin{align}
    \delta y^{i}(t) = L_{ij} \delta y^{i}(0),
\end{align}
where $L_{ij}(t)$ is the evolution matrix satisfying $\dot L_{ij}(t)= K_{il} L_{lj}$ and $L_{ij}(0)= \delta_{ij}$. The Lyapunov exponent measures the average exponential rate at which two nearby trajectories in a dynamical system diverge. It indicates the typical rate at which nearby orbits in the phase space either contract or expand. The principal Lyapunov exponent is determined by the eigenvalues of the matrix \(L_{ij}\), and is given by:
 \begin{eqnarray}
  \lambda= \lim_{t \xrightarrow{} \infty} \frac{1}{t} log\left(\frac{L_{ij}(t)}{L_{ij}(0)} \right).
 \end{eqnarray}
 Typically, the eigenvalues of the Jacobi matrix yield the Lyapunov exponent. A positive Lyapunov exponent indicates the presence of chaos in the system.

 In this section, we consider the circular motion of chargeless but massive particle around the black brane. We restrict ourselves to the equatorial orbit $(\phi_{1}= \frac{\pi}{2})$ circulating in the black p-brane with $x_{i} = \textit{const}$ and $\phi_{2} \equiv \varphi$, $\phi_{3}=\phi_{4}=..\phi_{8-p}=\textit{const}$, thereby reducing the Lagrangian to \cite{chandrasekhar1985mathematical} 
 \begin{flalign}\label{pt-Lag}
      2 \mathcal{L}= -\Delta_{+}\Delta_{-}^{-1/2} \dot t^{2}+ \Delta_{+}^{-1}\Delta_{-}^{\gamma} \dot r^{2}+r^{2} \Delta_{-}^{\gamma +1} \dot \varphi^2
 \end{flalign}
 where the dot represents the derivative with respect to proper time.\\
 Since the Lagrangian \ref{pt-Lag} is independent of the $t, ~\varphi$ therefore the corresponding generalised momenta $p_{t}$ and $p_{\varphi}$
 \begin{flalign}
     p_{t}= \frac{\partial \mathcal{L}}{\partial \dot t} \equiv E, ~~~
    p_{\varphi}= \frac{\partial \mathcal{L}}{\partial \dot \varphi} \equiv l
 \end{flalign}
 are conserved which are given by
 \begin{flalign}
    E&= - \Delta_{+}\Delta_{-}^{-1/2} \dot t \\
     l&= r^{2} \Delta_{-}^{\gamma +1} \dot \varphi
 \end{flalign}
 and the generalised momentum along r-coordinate is given by $p_{r}= \Delta_{+}^{-1}\Delta_{-}^{\gamma}\dot r$. We shall use the $p_{\varphi}$ and $l$ interchangeably meaning the same thing.\\
 Therefore the Hamiltonian becomes
 \begin{flalign}
     2 \mathcal{H} &= 2 \left(p_{t} \dot t + p_{\varphi} \dot + p_{r} \dot r - \mathcal{L} \right) \\
     2 \mathcal{H}&=  -\Delta_{+}\Delta_{-}^{-1/2} \dot t^{2}+ \Delta_{+}^{-1}\Delta_{-}^{\gamma} \dot r^{2}+r^{2} \Delta_{-}^{\gamma +1} \dot \varphi^2 \\
     2 \mathcal{H}&=  -\frac{p_{t}^2}{\Delta_{+}\Delta_{-}^{-1/2}}+ \frac{p_{\varphi}^2}{r^{2} \Delta_{-}^{\gamma +1}} + \frac{p_{r}^2}{\Delta_{+}^{-1}\Delta_{-}^{\gamma}}
 \end{flalign}
 From the Hamiltonian, the equations of motion are obtained as
 \begin{align}
     \dot t&= -\frac{p_{t} \sqrt{\Delta_{-}}}{\Delta_{+}}\\
     \dot p_{t}&= 0\\
     \dot r=& p_{r} \Delta_{+} \Delta_{-}^{-\gamma} \\
    \dot p_{r}&= \frac{1}{2} \Bigg( \gamma  \Delta _+ p_{r}^2 \Delta _-^{-\gamma -1} \Delta _-'-p_{r}^2 \Delta _-^{-\gamma } \Delta _+'+\frac{p_{t}^2 \Delta _-'}{2 \sqrt{\Delta _-} \Delta _+}-\frac{\sqrt{\Delta _-} p_{t}^2 \Delta _+'}{\Delta _+^2} \nonumber \\& +\frac{2 p_{\phi}^2 \Delta _-^{-\gamma -1}}{r^3}-\frac{(-\gamma -1) p_{\phi}^2 \Delta _-^{-\gamma -2} \Delta _-'}{r^2} \Bigg) \\
     \dot \varphi &= p_{\varphi} \frac{\Delta_{-}^{-1- \gamma}}{r^2} \\
     \dot p_{\varphi}&= 0
 \end{align}
 where $\prime$ represents the derivative with respect to $r$. 
 Next, we define
 \begin{align}
     F_{1} \equiv \frac{\dot r}{\dot t} &= -\frac{\Delta _+^2 \Delta _-^{-\gamma -\frac{1}{2}} p_r}{p_t}\\
     F_{2} \equiv \frac{\dot p_{r}}{\dot t} &= -\frac{\Delta _+ \Delta _-^{-\gamma -\frac{3}{2}} p_{\varphi }^2}{r^3 p_t}-\frac{(\gamma +1) \Delta _+ \Delta _-^{-\gamma -\frac{5}{2}} \Delta _{-}^{\prime} p_{\varphi }^2}{2 r^2 p_t}+\frac{\Delta _+ \Delta _-^{-\gamma -\frac{1}{2}} \Delta _{+}^{\prime} p_r^2}{2 p_t} \nonumber \\& 
     -\frac{\gamma  \Delta _+^2 \Delta _-^{-\gamma -\frac{3}{2}} \Delta _{-}^{\prime} p_r^2}{2 p_t}+\frac{\Delta _{+}^{\prime} p_t}{2 \Delta _+}-\frac{\Delta _{-}^{\prime} p_t}{4 \Delta _-}
 \end{align}
 Using the normalization condition for the four-velocity of a massive particle, $g_{\mu \nu} \dot x^{\mu} \dot x^{\nu}=-1$, we obtain
 \begin{align}
    \frac{\Delta _-^{-\gamma -1} p_{\varphi }^2}{r^2}+\Delta _+ \Delta _-^{-\gamma } p_r^2-\frac{\sqrt{\Delta _-} p_t^2}{\Delta _+}=-1
 \end{align}
 We substitute this expression into the $F_{1}$ and $F_{2}$ to eliminate $p_{t}$, we get
 \begin{align}
     F_{1}&= -\frac{\Delta_{+}^{3/2} r \Delta _{-}^{\frac{1}{4}-\frac{\gamma }{2}} p_r}{\sqrt{\Delta _- r^2 \left(\Delta _-^{\gamma }+\Delta _+ p_r^2\right)+p_{\varphi }^2}} \\
     F_{2}&= \frac{1}{4 \sqrt{\Delta _{+}} r^2 \sqrt{\Delta _{-} r^2 \left(\Delta _{-}^{\gamma }+\Delta _{+} p_r^2\right)+p_{\varphi }^2}} \Bigg[\Delta _{-}^{-\frac{\gamma }{2}-\frac{7}{4}} \Big(-4 \Delta _- \Delta _+ p_{\varphi }^2 
     \nonumber \\&
     - \Delta _{+} r \Delta _{-}^{\prime} \left((2 \gamma +3) p_{\varphi }^2+\Delta _{-} r^2 \left(\Delta _-^{\gamma }+(2 \gamma +1) \Delta _{+} p_r^2\right)\right) \nonumber \\&
     +2 \Delta _{-} r \Delta _{+}^{\prime} \left(\Delta_{-} r^2 \left(\Delta_{-}^{\gamma }+2 \Delta _{+} p_{r}^2\right)+p_{\varphi }^2\right)\Big) \Bigg]
\end{align}
Next we calculate the elements of Jacobian matrix $K_{ij}$ in the phase space $(r, p_{r})$ defined as:
\begin{equation*}
    K_{11} = \frac{\partial F_{1}}{\partial r},  K_{12} = \frac{\partial F_{1}}{\partial p_{r}}, K_{21} = \frac{\partial F_{2}}{\partial r},  K_{22} = \frac{\partial F_{2}}{\partial p_{r}}
\end{equation*}
Using the condition $p_{r}=\dot p_{r}$ = 0, we get the radial equilibrium orbit $r=r_{0}$ and thereof calculating the eigenvalues of $K_{ij}$, we obtain the following Lyapunov exponent
\begin{align} \label{lambda}
    \lambda^{2}&= -\frac{1}{16 r^2 \left(p_{\varphi }^2+r^2 \Delta _{-}^{\gamma +1}\right)^2} \Bigg[\Delta _{-}^{-\gamma -\frac{5}{2}} \Big((8 \gamma^2 + 14 \gamma + 30) \Delta _{+}^2 r^4 \Delta _-^{\gamma +1} \Delta _-'{}^2 p_{\varphi }^2 \nonumber \\
    &\quad -8 \Delta _{+} r^3 \Delta _{-}^{\gamma +2} p_{\varphi }^2 \left[(\gamma +2) r \Delta _+'\Delta _-'-(4 \gamma +5) \Delta _+\Delta _-'+(\gamma +2) \Delta _+ r \Delta _-''\right] \nonumber \\
    &\quad +8 r^2 \Delta _-^{\gamma +3} p_{\varphi }^2 \left[6 \Delta _+^2-r^2 \Delta _+'{}^2+2 \Delta _+ r \left(r \Delta _+''-\Delta _+'\right)\right] \nonumber \\
    &\quad +(2 \gamma +3) (2 \gamma +7) \Delta _+^2 r^2 \Delta _-'{}^2 p_{\varphi }^4+4 \Delta _-^2 p_{\varphi }^4 \{8 \Delta _+^2-r^2 \Delta _+'{}^2 \nonumber \\
    &\quad +2 \Delta _+ r \left(r \Delta _+''-2 \Delta _+'\right)\}-4 (2 \gamma +3) \Delta _+ \Delta _- r p_{\varphi }^4 \left[\Delta _-' \left(r \Delta _+'-2 \Delta _+\right)+\Delta _+ r \Delta _-''\right] \nonumber \\
    &\quad -4 \Delta _+ r^6 \Delta _-^{2 \gamma +3} \left(\Delta _-' \Delta _+'+\Delta _+ \Delta _-''\right)-4 r^6 \Delta _-^{2 \gamma +4} \left(\Delta _+'{}^2-2 \Delta _+ \Delta _+''\right) \nonumber \\
    &\quad +5 \Delta _+^2 r^6 \Delta _-^{2 \gamma +2} \Delta _-'{}^2\Big) \Bigg]
\end{align}
Note that the expression of the Lyapunov exponent involves both the charge of the p-brane and the angular momentum of the particle. The numerical values of $r_{0}$ are obtained by equating $p_{r}= F_{2}=0$. Table \ref{tab:ang} shows the values of the $r_{0}$ at specific parameters, without loss of generality for $p=3$. The table shows that $r_{0}$ decreases slightly with the increase of $l$. For more details on the numerics, see Appendix. Note that we cannot arbitrarily take any large value of $l$ as this may give rise to the complex root of the equation $F_{2} = 0$, which is true irrespective of p. 

\begin{table}[h!]
\begin{center}

\begin{tabular}{|c|c|c|c|c|c|c|} 
\hline
$l$ & 5 & 10 & 20 & 30 & 40 & 50\\
\hline
$r_{0}$ & 1.2873 & 1.2793 & 1.2772 & 1.2769 & 1.2768 & 1.2767\\
\hline
\end{tabular}
\caption{Positions of the circular orbits of the particle near 3-black brane at different values of $l$ when $Q=2.3$. The event (outer) horizon is located at $r_{+}= 1.05038$.}
\label{tab:ang}

\end{center}

\end{table}
Since the chaos bound can be rewritten as $\lambda^2 - \kappa^2  \le 0$, therefore the sign of $\kappa^2 - \lambda^2$ would decide the fate of the violation of the bound. With all the necessary ingredients, we now study the effect of $l$ and $Q$ on the behaviour of $\kappa^2 - \lambda^2$.
\begin{figure}[h!]
\begin{subfigure}
  \centering
  \includegraphics[width=0.41\textwidth]{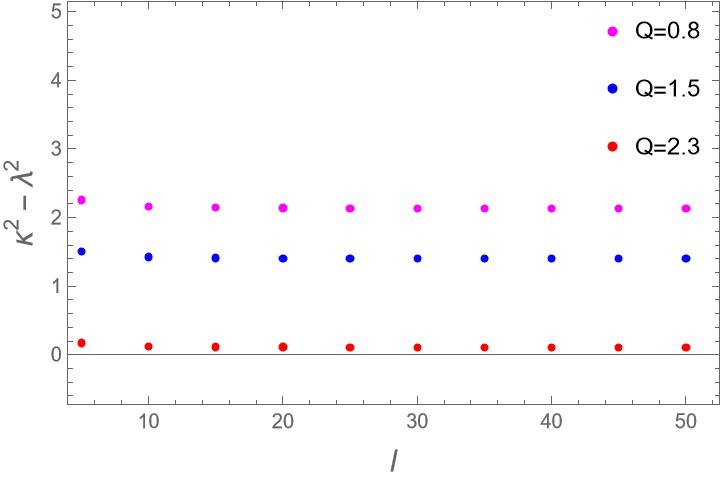}
\put(-175,125){(a)}
\end{subfigure}
\hspace{12mm}
\begin{subfigure}
  \centering
  \includegraphics[width=0.41\textwidth]{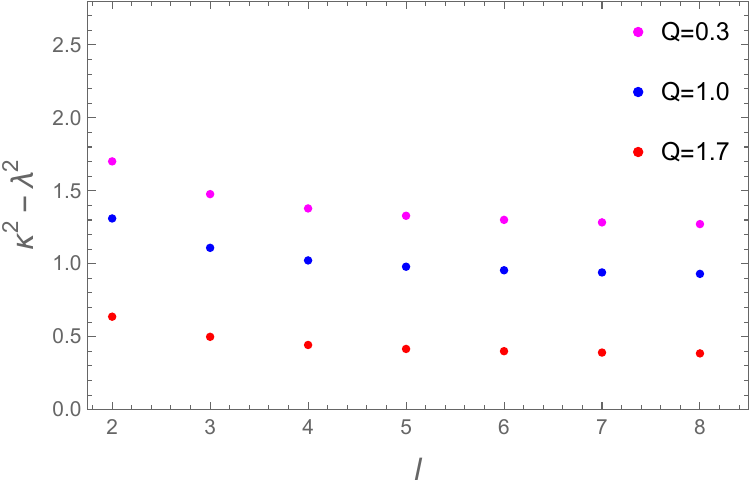}
\put(-175,125){(b)}
\end{subfigure}
\caption{Plots showing $\kappa^2 - \lambda^2$ as a function of $l$ for different Q. (a) $p=3$(left panel) and  (b) $p=4$(right panel).}
\label{point-p=3,4}
\end{figure}
\\
\begin{figure}[h!]
\begin{subfigure}
  \centering
  \includegraphics[width=0.41\textwidth]{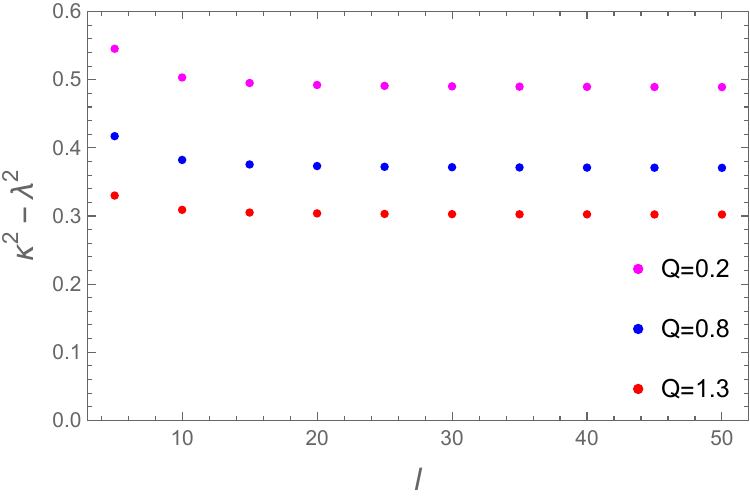}
\put(-175,125){(a)}
\end{subfigure}
\hspace{12mm}
\begin{subfigure}
  \centering
  \includegraphics[width=0.41\textwidth]{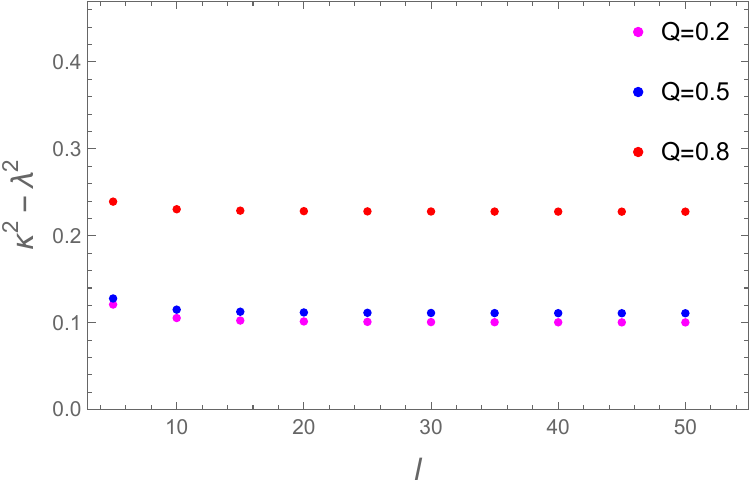}
\put(-175,125){(b)}
\end{subfigure}
\caption{Plots showing $\kappa^2 - \lambda^2$ as a function of $l$ for different Q. (a) $p=5$(left panel) and  (b) $p=6$(right panel).}
\label{point-p=5,6}
\end{figure}
\newline
\begin{figure}
    \centering
    \includegraphics[width=0.41\textwidth]{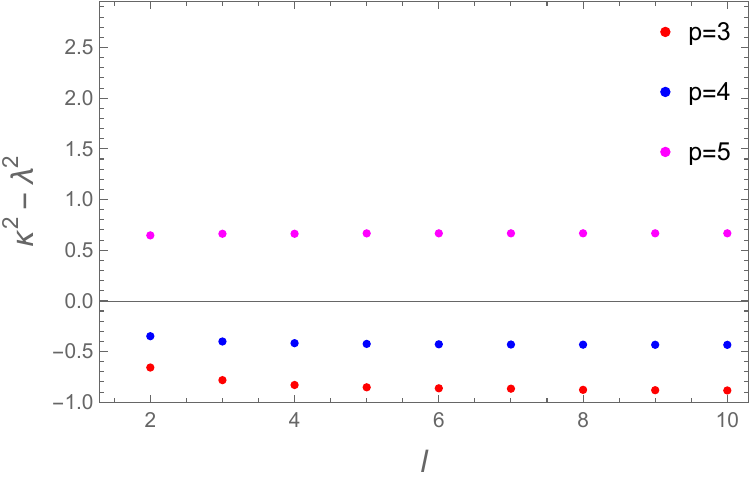}
    \caption{Plot showing $\kappa^2 - \lambda^2$ as a function of $l$ corresponding to extremal p-brane.}
    \label{fig:extremal-particle}
\end{figure}
We highlight the key observations as follows:
\begin{itemize}
    \item On keeping $Q$ fixed, the value of $\kappa^2 - \lambda^2$  at first decreases with the increase of $l$ and effectively remains constant at large $l$. Since, $\kappa^2 - \lambda^2 > 0$ for all $l$, we conclude there is no violation of bound (Fig \ref{point-p=3,4}, Fig \ref{point-p=5,6})    
    \item The variation of charge (or Hawking temperature) of the p-brane brings interesting features to the Lyapunov exponent. For p = 3, 4 and 5 brane, the quantity $\kappa^2 - \lambda^2$ seems to decrease with the increase of charge at a fixed angular momentum $l$. Note that for p =3 and Q = 2.3, $\kappa^2 - \lambda^2$ approaches the horizontal line ($\kappa^2 = \lambda^2$) thus saturating the bound (fig \ref{point-p=3,4}(a)). However, for p = 6, quantity $\kappa^2 - \lambda^2$ increases with charge and the corresponding curve moves away from the horizontal line (fig \ref{point-p=5,6}(b)). This could be due to the increase of charge, the Hawking temperature decreases for p = 3, 4, 5 but increases for p = 6.
\end{itemize}
Next, we investigate the chaos bound in the extremal limit (fig \ref{fig:extremal-particle}). For our choice of parameters, this corresponds to Q = 2.5 (p = 3), Q = 2 (p = 4), Q = 1.5 (p = 5) and Q = 1 (p = 6). For the p = 3 and p = 4 brane, we see a clear violation of the bound as $\kappa^2 - \lambda^2 < 0$ and we do not observe significant variations with $l$. However, for p = 5 brane, all the points lie above  $\kappa^2 = \lambda^2 $ thus satisfying the strict inequality. We have not shown the p = 6 case as the extremal 6-brane corresponds to $T\rightarrow \infty$. Also, we have found that the equilibrium position, $r_{0} >> r_{+}$ for any $l$ which eventually leads to $\lambda$ = 0. Thus, in this case the bound is trivially followed.

Finally, from the eq \ref{kappa} and \ref{lambda}, it is easy to verify that the bound saturates at the (outer) horizon: $\lambda = \kappa$, a common feature observed in many spherically symmetric and static backgrounds \cite{Hashimoto:2016dfz,xie2023circular}.

 \section{Closed string around black p-brane}\label{string probe}
  The dynamics of a closed circular string in an arbitrary curved background can be described using the Polyakov action given as,
 \begin{equation} \label{eqn: polyakov}
 S = -\frac{1}{2\pi \alpha^{\prime}} \int d\sigma d\tau \sqrt{-g} g^{\alpha \beta} G_{\mu \nu}(\partial_{\alpha} X^{\mu}\partial_{\beta} X^{\nu})
 \end{equation}
 where $\alpha^{\prime} = l_s^{2}$ ($l_s$ represents the string length). $X^{\mu}$ represents the target space co-ordinates, $G_{\mu \nu}$ is the target space metric and $g_{\alpha \beta}$ is the worldsheet metric. We choose the conformal gauge ($g^{\alpha\beta}$ = $\eta^{\alpha\beta}$) which leads to  vanishing of energy-momentum tensor $T_{\alpha \beta}$ = 0. Using this condition, we find
 \begin{flalign}
 G_{\mu\nu} \partial_{\tau}X^{\mu} \partial_{\sigma}X^{\nu}& = 0 \label{eqn: gauge_conformal}\\
G_{\mu\nu} \Big( \partial_{\tau}X^{\mu} \partial_{\tau}X^{\nu}+\partial_{\sigma}X^{\mu} \partial_{\sigma}X^{\nu} \Big)& = 0\label{eqn: gauge_conformal2}
\end{flalign}
The target space metric $G_{\mu \nu}$ is given by equation \ref{eqn: brane_metric2}.

Now we consider the following closed string ansatz for generic p-brane:
\begin{eqnarray}
    t = t(\tau) \ ,  r= r(\tau) \ , \phi_{1} = \phi_{1}(\tau) \ , \phi_{2} = n\sigma \ , \phi_{3} = ...=\phi_{8-p} = constant
\end{eqnarray}
where $n$ denotes the winding number of the string along $\phi_{2}$ direction. We assume the spatial coordinates $x^{i}$ are constant.

The corresponding Lagrangian and Hamiltonian are given by
\begin{flalign}
    L & = -\frac{1}{2\pi\alpha^{\prime}} \Big( \Delta_{+}\Delta_{-}^{-1/2} \dot t^{2}-\Delta_{+}^{-1}\Delta_{-}^{\gamma} \dot r^{2}-r^{2}\Delta_{-}^{\gamma +1} ( \dot \phi_{1}^{2}- n^{2} \sin^{2}\phi_1)) \\
    H & = \frac{\pi \alpha^{\prime}}{2} \Big( \Delta_{+}\Delta_{-}^{-\gamma} p_{r}^{2}+\frac{p_{\phi_{1}^{2}}}{r^2 \Delta_{-}^{\gamma +1} } - \frac{p_{t}^{2}}{\Delta_{+}\Delta_{-}^{-1/2}} \Big) + \frac{1}{2 \pi \alpha^{\prime}}n^{2}\Delta_{-}^{\gamma +1}r^{2}\sin^{2}\phi_{1} \label{H}
\end{flalign}
The equations of motion can be worked out as follows:
\begin{flalign} 
\dot{p_{t}} & = 0 \label{eqn: energy2}\\ 
\dot{t} & = -\pi \alpha^{\prime}\Delta_{-}^{1/2}
\Delta_{+}^{-1} p_{t}\\
\dot{p_{r}} & = \frac{\pi \alpha^{\prime}}{2}\frac{\partial} {\partial r} \Big( -p_{r}^{2}\Delta_{-}^{-\gamma}\Delta_{+}+p_{t}^{2}
\Delta_{-}^{1/2}\Delta_{+}^{-1} - p_{\phi_{1}^{2}} \frac{1}{r^2 \Delta_{-}^{\gamma +1}}\Big)- \frac{n^{2}}{2\pi \alpha^{\prime}} \frac{\partial }{\partial r}(r^2 \Delta_{-}^{\gamma +1} \sin^{2}\phi_{1} \Big)\\  
\dot{r} & = \pi \alpha^{\prime} \Delta_{+}\Delta_{-}^{-\gamma} p_{r} \\
\dot{p}_{\phi_{1}} &  = - \frac{n^{2}}{\pi \alpha^{\prime}} r^{2}\Delta_{-}^{\gamma +1} \sin\phi_{1}\cos\phi_{1}\\
\dot{\phi_{1}} & = \pi \alpha^{\prime} \frac{p_{\phi_{1}}}{r^2 \Delta_{-}^{\gamma +1}}\label{eqn: phieqn}  
\end{flalign}
\newline
Since, t is a cyclic coordinate, $p_{t} = E$ (constant of motion). The conformal gauge constraint gives  H = 0. The Hamiltonian has some interesting features. The Hamiltonian \ref{H} can be seen as the sum of the kinetic energy term and the effective potential term (for $r$ and $\phi$). If we consider the radial motion of the system, it possesses the effective attractive potential which has the divergence at the horizon whereas the $\phi$- dependent part contains two repulsive terms. The string exhibits different modes as a function of $r$ for specific values of $p$ \cite{basu_chaotic_2017,PandoZayas:2010xpn,frolov_chaotic_1999}. For the sake of completeness, we have also analysed the fixed points of our system which are provided in the Appendix. 

\subsection{Classical string solution and non-integrability}\label{non-integrability}
 

 To study the non-integrability in the p-brane it is convenient to use the metric \ref{eqn: cuvrobic}. This gives the following Lagrangian
\begin{equation}
    \mathcal{L} = -\frac{1}{2\pi\alpha^{\prime}} \Big( \mathcal{F}\dot{t}^{2}- \frac{\dot{r}^{2}}{\mathcal{F}}-r^{2} \sqrt{H}(\dot{\phi_{1}}^{2}-n^{2}\sin^{2}\phi_{1}) \Big)
\end{equation}
The corresponding Hamiltonian is given by
\begin{eqnarray}
    \mathcal{H}= -\frac{ p_{t}^2}{2 \mathcal{F} } + \frac{p_{r}^2 \mathcal{F}}{2} + \frac{p_{\phi_{1}}^2}{2 r^2 \sqrt{H}} + \frac{n^2 r^2 \sqrt{H} \sin^2{\phi_{1}}}{2}
\end{eqnarray}
with the equations of motion
\begin{flalign}
    \mathcal{F}\dot{t} = E = constant 
\end{flalign}
\begin{eqnarray}
   \ddot r \mathcal{F} - \frac{1}{2} \mathcal{F}^\prime \dot r ^2 + \frac{\mathcal{F}^2 \mathcal{F}^\prime}{2} \dot t^2  - \mathcal{F}^2 \Big(\dot \phi_{1} ^2 - n^2 \sin^2 \phi_{1} \Big) \Big(r\sqrt{H} + \frac{r^2 H^\prime}{4 \sqrt{H}} \Big) &=0 \\
   \ddot \phi_{1} + \Big( \frac{2}{r} + \frac{H^\prime}{2 H} \Big) \dot r \dot \phi_{1} + \frac{n^2 \sin2\phi_{1}}{2}& =0
\end{eqnarray}
Now, corresponding to the closed string in the background, we consider an invariant plane solution $(r,p_{r}, \phi_1, p_{\phi_1})= (\overline{r}, \frac{E}{\mathcal{F}(\overline{r})}, N \pi, 0)$ where $N \in  \mathbb{Z}$ and $\overline{r} \equiv r_{0}+ E \tau$ with $r_{0}$=constant. It is an easy exercise to verify that this plane satisfies the constraint  \ref{eqn: gauge_conformal2}. (Here we have only taken the positive solution corresponding to the $\dot r^2= E^2$ since the $r(\tau)$ is chosen to be non-negative with the arbitrary choice of $E$). We expand the equation of motion for $\phi_1$ using 
$$\phi_{1}(\tau) = N \pi + \eta(\tau), ~~ |\eta| << 1$$ 
then up to the first order in $\eta$, the corresponding NVE is obtained as follows:
\begin{flalign}
     \ddot \eta + \Bigg( \frac{2}{\overline{r}} + \frac{H'(\overline{r})}{2 H(\overline{r})} \Bigg) E \dot \eta + n^2 \eta =0
\end{flalign}
The non-integrability would be determined by the existence of the Liouville solution on applying the Kovacic algorithm\footnote{We have used Maple in-built function kovacicsols.}. The Kovacic algorithm fails; therefore, the above NVE does not possess the Liouville solution. However, integrability is restored for particle motion $(n = 0)$ where the NVE admits the Liouviilian solution which is expected for the particle scenario as shown earlier in \cite{Dutta:2023yhx,stepanchuk_nonintegrability_2013}. Thus, the closed string breaks the integrability structure even in the non-extremal black p-brane. In the next section, we will gather sufficient evidence for the chaos in the p-brane followed by the numerical investigation of the chaos bound.\\

\subsection{Lyapunov Exponent and Chaos Bound
}\label{numerical bound}
 
 \begin{table}[h] \label{table- LE}
\begin{center}
    \begin{tabular}{| *{7}{c|} }
    \hline
p-values   & \multicolumn{1}{c|}{$r(0)$} 
& \multicolumn{1}{c|}{$pr(0)$}
& \multicolumn{1}{c|}{$E$} 
& \multicolumn{1}{c|}{$Q$} 
& \multicolumn{1}{c|}{$\lambda$} 
                 & \multicolumn{1}{c|}{Mode}     \\
    \hline
3   &  10  & 0.02   & 5 & 2.2  &  0.0231  & Capture\\
   &   10  & 0.02   &  2   & 2.2 & 0.0008 & Escape  \\
    \hline
4   &  12     & 0.02      &  2  & 1.7 & 0.0153 & Capture     \\
       &   12  & 0.04   &  1   & 1.7 & 0.0007 &  Escape \\
    \hline
5  
   &   12  & 0.1   & 2    & 1.2 & 0.0272 &  Capture \\
   &   12 &   0.1 & 1 &  1.2 & 0.0008   &  Escape\\
   \hline
6   &   13    &  0     &  2  & 0.7  &  0.086 & Capture  \\
    &   13  & 0.4 & 1 &  0.7 & 0.0009 &  Escape\\
    \hline
\end{tabular}
\caption{The table represents the different dynamics of the string at various parameters for n=1 along with the value of the largest Lyapunov exponent $\lambda$. }
\label{tab:mytable}
\end{center}
\end{table}
To observe the existence of chaos, we compute the Lyapunov exponent for different p-values. The initial conditions and the control parameters are chosen in such a way that the string quickly escapes from the horizon. In this scenario, we observe that the maximum or Largest Lyapunov Exponent (LLE) are 0.00143, 0.001, 0.00103 and 0.0011 for p =3,4,5 and 6 (fig \ref{fig: LLE-p=3,4}(a),(b), \ref{fig:LLE-p=5,6}(a),(b)) respectively. The plots show that the dynamical system exhibits chaos even for generic p-values, $p<7$.  Next, we choose the initial conditions and the control parameters in such a way that the string does not fly away from the horizon during the evolution. Details on the different string trajectories are provided in the appendix. The corresponding maximum Lyapunov exponents are 0.0099, 0.0116, 0.0183, 0.0218 in fig \ref{fig: LLE-p=3,4(near)}(a),(b),\ref{fig: LLE-p=5,6(near)}(a),(b) respectively. This shows that the chaotic behaviour is much more prominent in the latter scenario. Note that in these cases the string shows oscillations in the beginning and eventually gets captured in the horizon. This variation of LLE between two scenarios reflects the dependency of LLE on $r(0)$, $T$ and $n$. The nature of the string dynamics corresponding to different parameters is listed in Table \ref{tab:mytable}.. 


\begin{figure}[h!]
\begin{subfigure}
  \centering
  \includegraphics[width=0.41\textwidth]{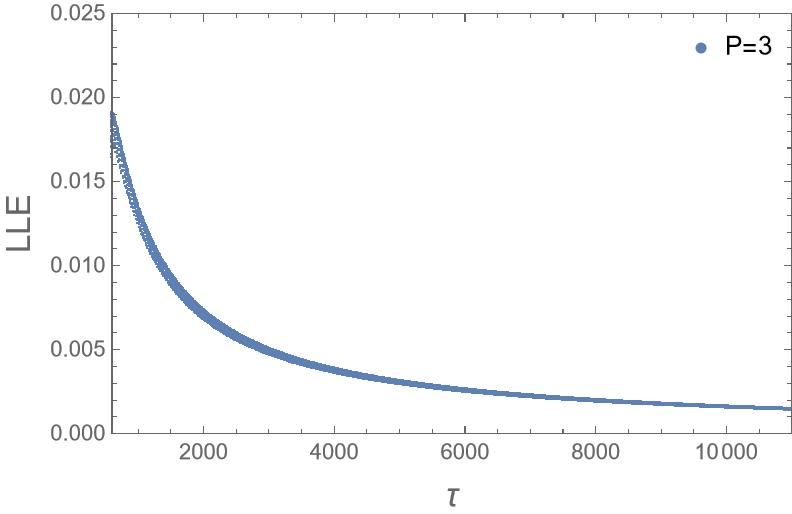}
\put(-180,120){(a)} 
\end{subfigure}
\hspace{12mm}
\begin{subfigure}
  \centering
  \includegraphics[width=0.41\textwidth]{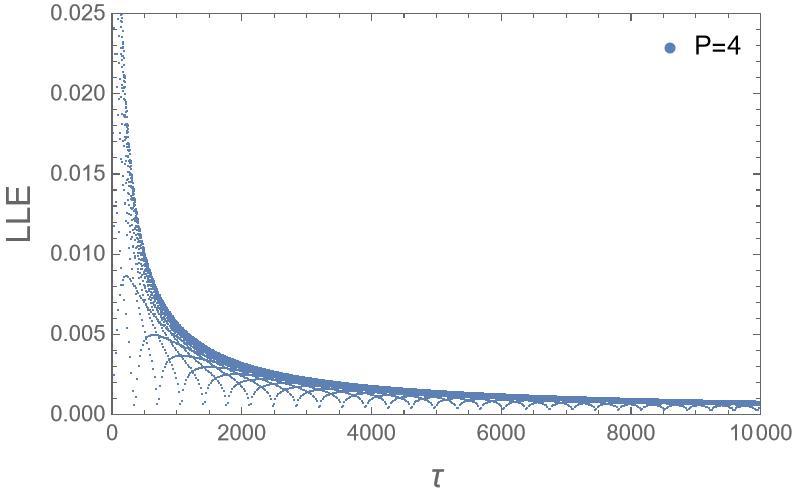}
\put(-180,120) {(b)} 
\end{subfigure}
\caption{Largest Lyapunov Exponent: (a) $p=3$, $r(0) = 15$, $E=4$, $p_{r}(0) = 0$,$\phi_{1}(0)=0$, Q = 0.5 and (b) $p=4$, $r(0) = 15$, $E=5$, $p_{r}(0) = 1$, $\phi_{1}(0)=0$, Q = 0.01.}
\label{fig: LLE-p=3,4}
\end{figure}
\begin{figure}[h!]
\begin{subfigure}
  \centering
  \includegraphics[width=0.41\textwidth]{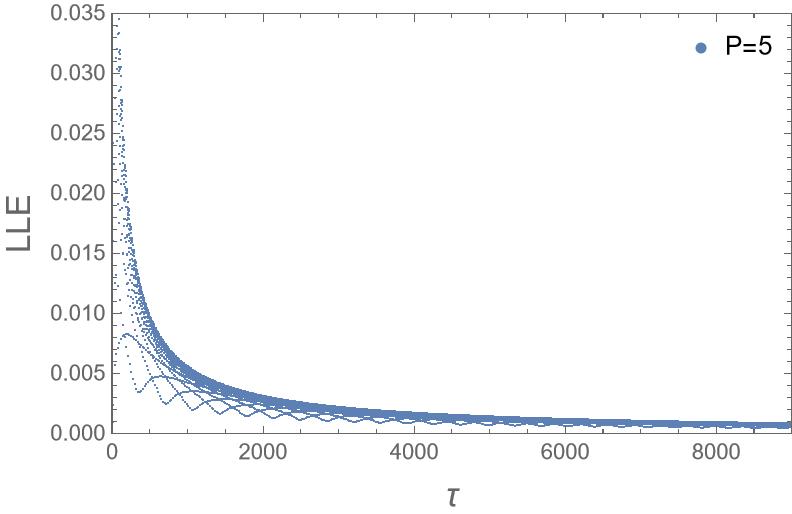}
\put(-180,120){(a)} 
\end{subfigure}
\hspace{12mm}
\begin{subfigure}
  \centering
  \includegraphics[width=0.41\textwidth]{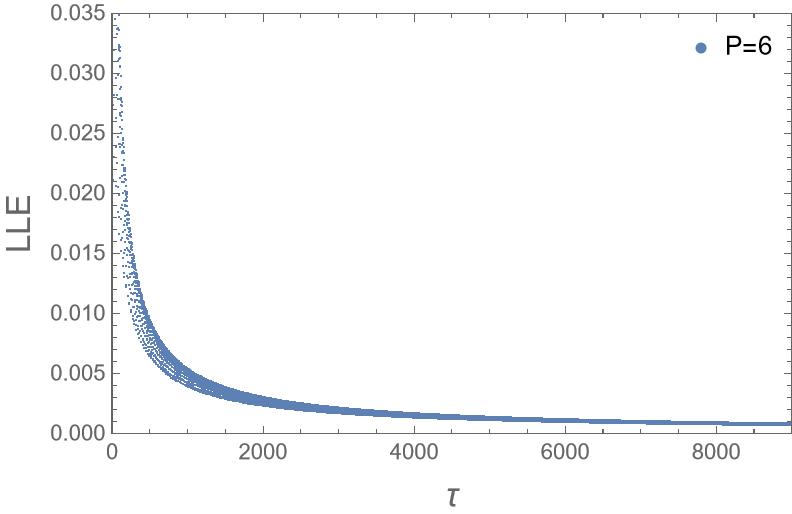}
\put(-180,120){(b)} 
\end{subfigure}
\caption{Largest Lyapunov Exponent: (a) $p=5$, $r(0) = 15$, $E=5$, $p_{r}(0) = 1$,$\phi_{1}(0)=0$, Q = 1 and (b) $p=6$, $r(0) = 15$, $E=5$, $p_{r}(0) = 2$, $\phi_{1}(0)=0$, Q = 0.01.}
\label{fig:LLE-p=5,6}
\end{figure}
\begin{figure}[h!]
\begin{subfigure}
  \centering
  \includegraphics[width=0.41\textwidth]{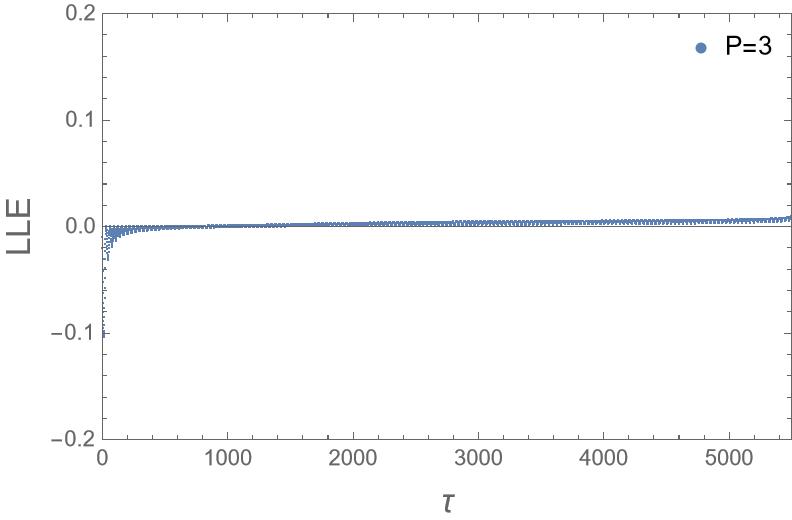}
\put(-180,120){(a)} 
\end{subfigure}
\hspace{12mm}
\begin{subfigure}
  \centering
  \includegraphics[width=0.41\textwidth]{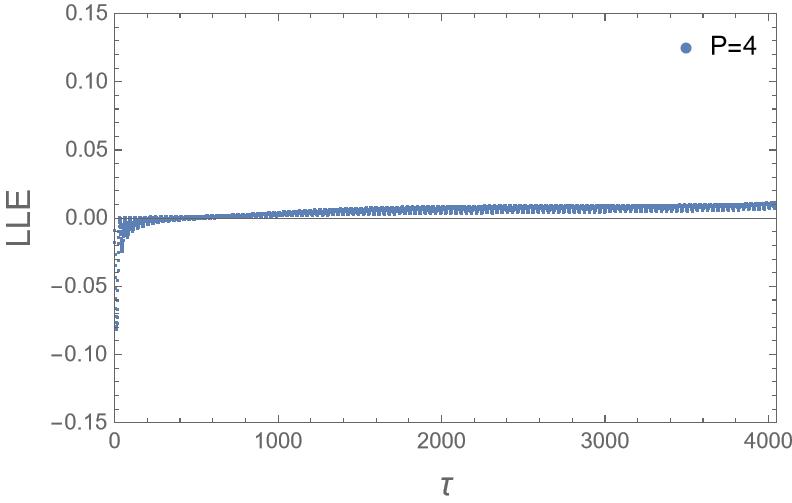}
\put(-180,120){(b)} 
\end{subfigure}
\caption{Largest Lyapunov Exponent: (a) $p=3$, $r(0) = 16$, $E=5$, $p_{r}(0) = 0$,$\phi_{1}(0)=0$, Q = 1 and (b) $p=4$, $r(0) = 18$, $E=5$, $p_{r}(0) = 0.03$, $\phi_{1}(0)=0$, Q = 1.}
\label{fig: LLE-p=3,4(near)}
\end{figure}
\begin{figure}[h!]
\begin{subfigure}
  \centering
  \includegraphics[width=0.41\textwidth]{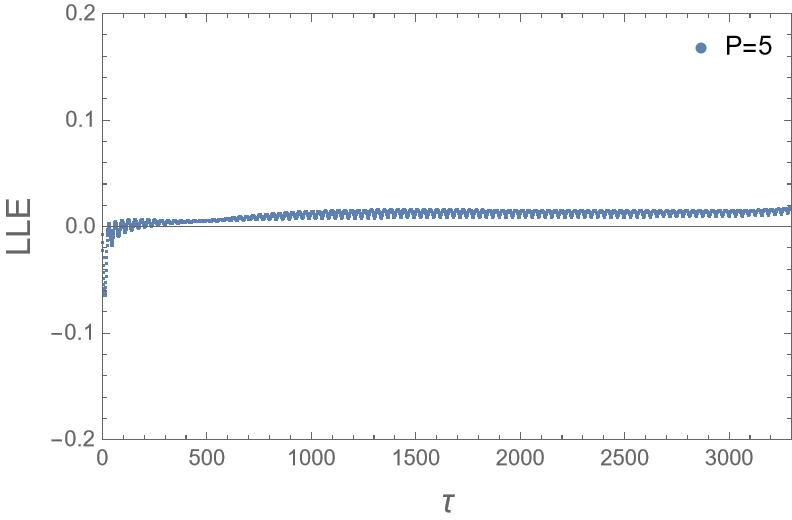}
\put(-180,120){(a)} 
\end{subfigure}
\hspace{12mm}
\begin{subfigure}
  \centering
  \includegraphics[width=0.41\textwidth]{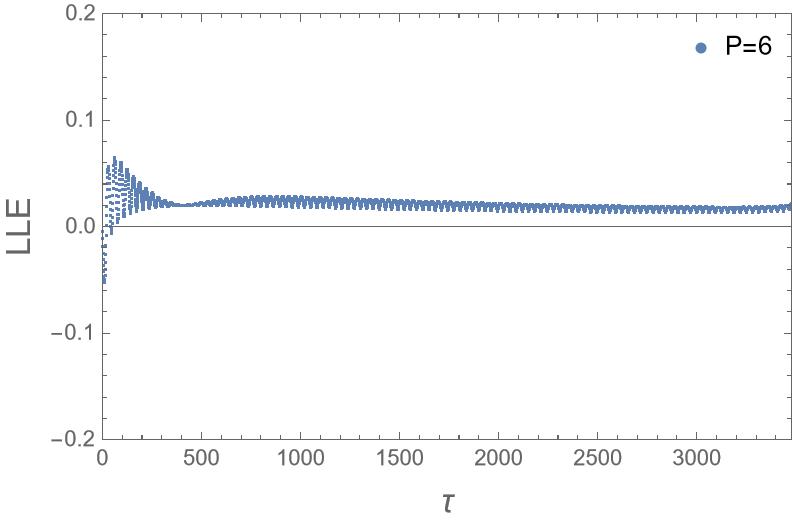}
\put(-180,120){(b)} 
\end{subfigure}
\caption{Largest Lyapunov Exponent: (a) $p=5$, $r(0) = 20$, $E=5$, $p_{r}(0) = 0.17$,$\phi_{1}(0)=0$, Q = 1 and (b) $p=6$, $r(0) = 20$, $E=5$, $p_{r}(0) = 1$, $\phi_{1}(0)=0$, Q = 0.3.}
\label{fig: LLE-p=5,6(near)}
\end{figure}
\begin{figure}[h!]
\begin{subfigure}
  \centering
  \includegraphics[width=0.41\textwidth]{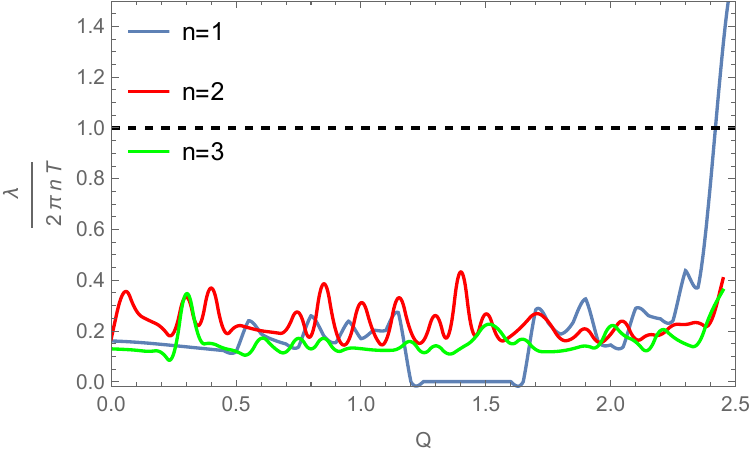}
\put(-175,110){(a)}
\end{subfigure}
\hspace{12mm}
\begin{subfigure}
  \centering
  \includegraphics[width=0.41\textwidth]{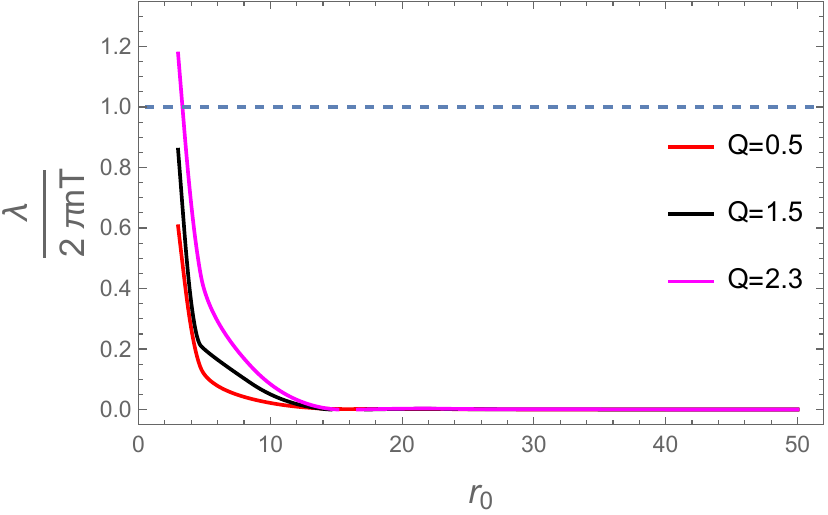}
\put(-175,110){(b)}
\end{subfigure}
\caption{Characteristics of p = 3. Plots showing the behaviour of  $\frac{\lambda}{2 n \pi T}$ for (a) variation of charge Q over the range [0,2.5] with $r(0)= 4.5$ and different values of $n=1,2,3$
and (b) variation of the $r(0)$ 
with different $Q=0.5,1.5,2.3$ and n = 1. For both plots, $E= 3.5,~~ p_{r}(0)=0, ~~ \phi_{1}(0)=0$.}

\label{fig: p=3(stringbound)}
\end{figure}
\begin{figure}[h!]
\begin{subfigure}
  \centering
  \includegraphics[width=0.41\textwidth]{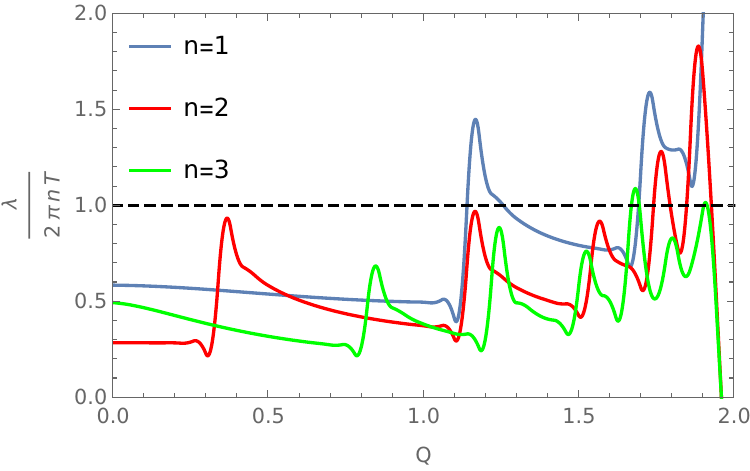}
\put(-175,110){(a)}
\end{subfigure}
\hspace{12mm}
\begin{subfigure}
  \centering
  \includegraphics[width=0.41\textwidth]{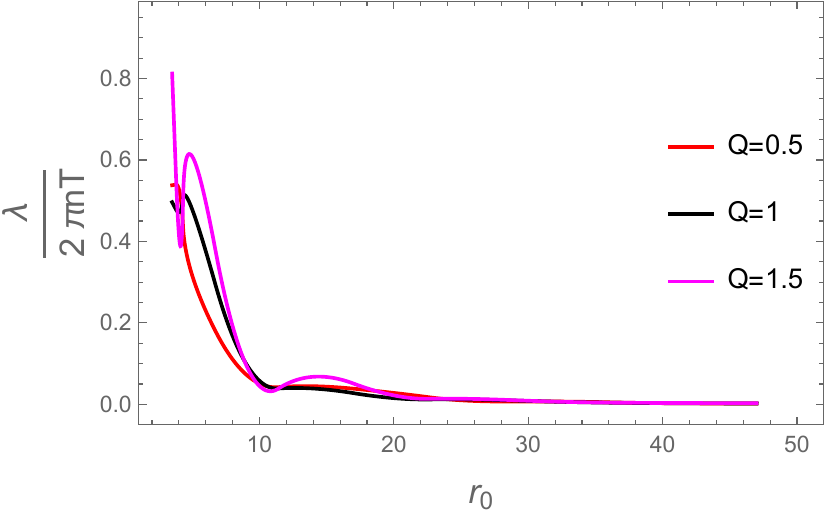}
\put(-175,110){(b)}
\end{subfigure}
\caption{Characteristics of p = 4. Plots showing the behaviour of $\frac{\lambda}{2 n \pi T}$ for (a) variation of charge Q over the range [0,2]
and $r(0) = 3.5$ with different $n=1,2,3$ and (b) variation of $r(0)$
  with different $Q=0.5,1,1.5$ and $n=1$. For both the plots, $E= 5.2,~~ p_{r}(0)=0, ~~ \phi_{1}(0)=0$.
}
\label{fig:p=4(stringbound)}
\end{figure}
\begin{figure}[h!]
\begin{subfigure}
  \centering
  \includegraphics[width=0.41\textwidth]{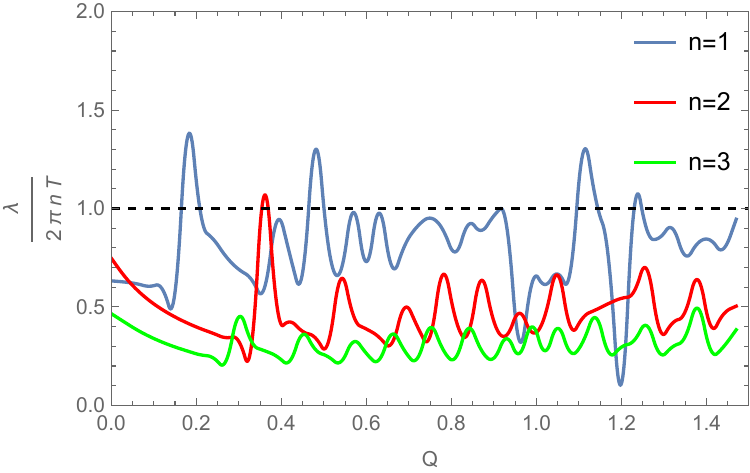}
\put(-175,110){(a)}
\end{subfigure}
\hspace{12mm}
\begin{subfigure}
  \centering
  \includegraphics[width=0.41\textwidth]{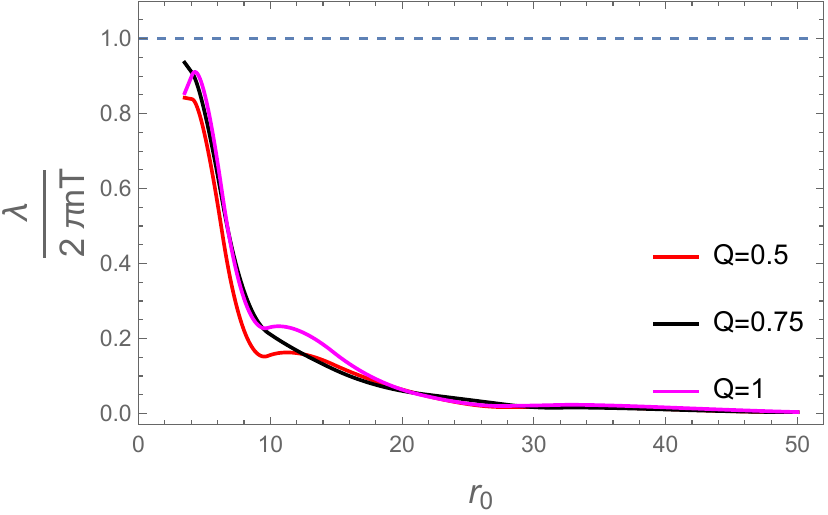}
\put(-175,110){(b)}
\end{subfigure}
\caption{Characteristics of p = 5. Plots showing the behaviour of $\frac{\lambda}{2 n \pi T}$ for (a) variation of Q over [0,1.5]
and $r(0) = 4.5$ with different $n=1,2,3$ and (b) over $r(0)$
  with different $Q=0.5,0.75,1$  and $n=1$. For both the plots, $E= 2.3,~~ p_{r}(0)=0.01, ~~ \phi_{1}(0)=0$.}
\label{fig:p=5(stringbound)}
\end{figure}
\begin{figure}[h!]
\begin{subfigure}
  \centering
  \includegraphics[width=0.41\textwidth]{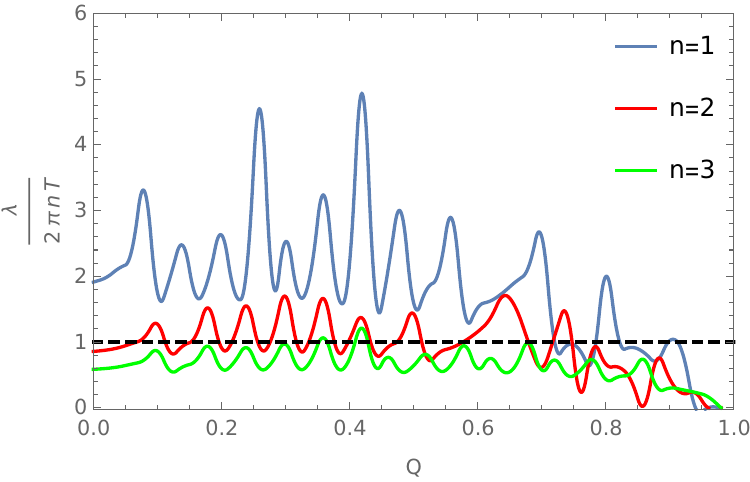}
\put(-175,110){(a)}
\end{subfigure}
\hspace{12mm}
\begin{subfigure}
  \centering
  \includegraphics[width=0.41\textwidth]{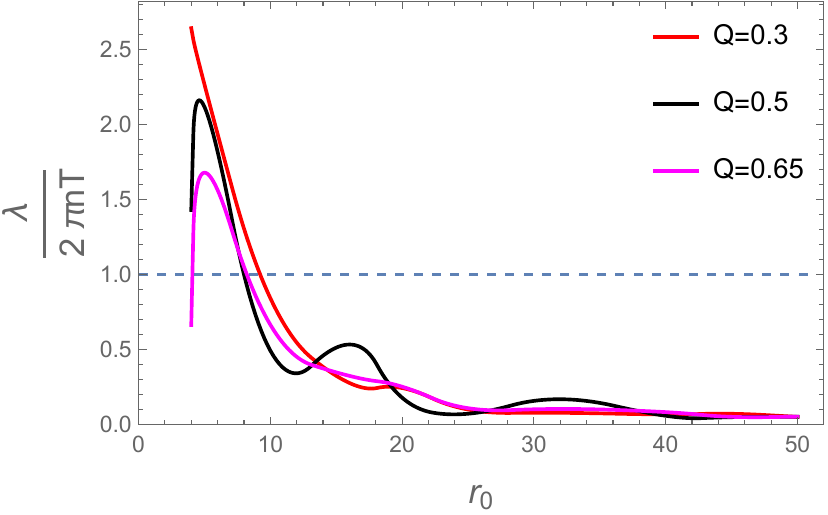}
\put(-175,110){(b)}
\end{subfigure}
\caption{Characteristics of p = 6. Plots showing the behaviour of $\frac{\lambda}{2 n \pi T}$ for (a) range of Q over [0,1]
and $r(0) = 5$ with different $n=1,2,3$ and (b) for variation of $r(0)$ 
  with different $Q=0.3,0.5,0.65$ and and $n=1$. For both the plots, $E= 2.3,~~ p_{r}(0)=0.01, ~~ \phi_{1}(0)=0$.}
\label{fig:p=6(stringbound)}
\end{figure}
\begin{figure}[h!]
    \centering
    \includegraphics[width=0.5\textwidth]{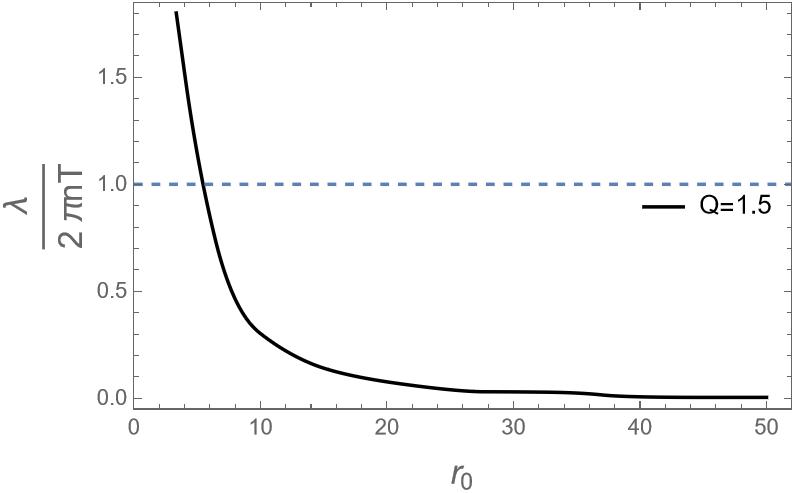}
    \caption{Characteristics of extremal p = 5 brane. Plots showing the variation of $\frac{\lambda}{2 n \pi T}$ for different $r(0)$. The other parameters are the same as in Fig 10(b).}
    \label{fig:p=5(extremal)}
\end{figure}

 Next, we determine whether the aforementioned generalised MSS inequality remains valid in the black p-brane background. We plot the ratio $\frac{\lambda}{2 \pi n T}$ at different initial conditions $r(0)\equiv r_0$ and at different charges $Q$ in Fig \ref{fig: p=3(stringbound)}-\ref{fig:p=6(stringbound)}. The various observations are summed up as follows: 
\begin{itemize}
    \item 
    In most cases, every p-brane satisfies the generalised inequality, but in a few cases, especially when the string is initially close to the horizon, we might observe the bound violation. However, as we move away from the horizon of the black brane, the inequality gets stronger with smaller values of the Lyapunov exponent, Fig \ref{fig: p=3(stringbound)}-\ref{fig:p=6(stringbound)} (right panel). Thus, the influence of charge is negligible when the string is initially far away from the horizon. Similar observations have been confirmed earlier in the works \cite{Dutta:2023yhx}.
    
    \item  The effect of charge and the winding number on chaotic dynamics is complex, especially in the near-horizon scenarios. The quantity $\frac{\lambda}{2 \pi n T}$ changes non-monotonically as a function of Q and n. However, as we scan over Q, the bound violation is significant for n = 1 (fig 10(a), 11(a)). This is not be true for p = 3 and p = 4 (fig 8(a), fig 9(a)) where the quantity $\frac{\lambda}{2 \pi n T}$ crosses unity as we reach closer to the extremal value of Q. 
\end{itemize}
    Lastly, we discuss the bound on extremal p brane. Note that for the extremal p=3 and p=4, $2\pi T \rightarrow 0$ and we have numerically tested (not shown)  that $\lambda\sim 10^{-3}$ for large $r(0)$ and $\lambda \approx 1$ for $2.85~ r_{+} < r(0) <3.20~r_{+}$\footnote{Due to the numerical stiffness of the system, we cannot approach the horizon $r_{+}$ exactly.}. This also demonstrates the bound violation is much more significant when the string is initially close to the horizon. Similarly, for the extremal 5-brane ($ 2 \pi T \rightarrow$ finite), the quantity $\frac{\lambda}{2 \pi n T}$ decreases with the increase of $r(0)$ again justifying the bound violation in the near horizon scenario (fig \ref{fig:p=5(extremal)}). However, for the extremal p=6 brane ($2 \pi T \rightarrow \infty$), $\frac{\lambda}{2 \pi n T}$ $\rightarrow 0$  which trivially shows no signature of bound violation. Thus, both circular geodesics and closed string satisfy the MSS inequality in the case of extremal 6-brane.



    

\section{Conclusions and Future Directions}\label{conclusions}
In this work, we have investigated the chaotic responses of time-like circular geodesics as well as a closed pulsating string around black p-brane and checked the universality of the chaos bound. Related to the universal bound, we conclude our results in Table \ref{tab:part} and \ref{tab:str}. For the case of circular geodesics, in the vicinity of the horizon, we find that the LLE is almost insensitive to the variations in angular momentum but sensitive to the charge of the black brane. In the non-extremal p-brane, the bound is satisfied. However, we observe bound violation in the extremal limit (p = 3 and p = 4) whereas, p = 5, 6 brane does not show any bound violation even in the extremal limit. In the case of a closed string, irrespective of p, the chaotic behaviour is noticeable when the string starts in the vicinity of the horizon. In this scenario, the inequality \ref{eqn: genbound} gets violated even for a higher winding number ($n>1$) by appropriately modulating the parameters. However, if the string starts far from the horizon, $LLE\rightarrow 0$ implies negligible chaos. Moreover, for the extremal p = 3,4 and 5 brane, we observe bound violation depending on the initial location of the string and once again, no violation is observed for the extremal p = 6 brane. The presence of chaos implies the analytical non-integrability of the closed string around the p-brane whereas one recovers the integrability for point particle scenario (n=0). Also, our studies can be attributed to the different scattering modes through the lens of the chaotic scattering amplitude \cite{frolov_chaotic_1999,Dutta:2023yhx} at the classical level. It would be interesting to explore the quantum chaos present in the system \cite{pando2013string}. Furthermore, through black hole/string correspondence, the chaos in string scattering could enhance our understanding of the quantum chaotic nature of black holes.

\begin{table}[h!]

\begin{center}

\begin{tabular}{|c|c|c|c|c|c|c|} 
\hline
$p \rightarrow$ & 3 &4 & 5 &6 \\
\hline
extremal &  \XSolidBrush & \XSolidBrush & \checkmark & \checkmark\\
\hline
non-extremal & \checkmark & \checkmark & \checkmark & \checkmark\\
\hline
\end{tabular}
\caption{Table for circular geodesics in the vicinity of the horizon. The symbol \checkmark (\XSolidBrush) represents the non-violation (violation) of the bound.}
\label{tab:part}

\end{center}

\end{table}

\begin{table}[!htb]
\small\sffamily\centering%
\setlength{\extrarowheight}{2.5pt}
\begin{tabularx}{\linewidth}{|l|*{5}{X|}c|}
\hline
$p \rightarrow$ & &  3  & 4& 5&6 \\
\hline
extremal & near-horizon  & \XSolidBrush & \XSolidBrush & \XSolidBrush & \checkmark\\
\cline{2-2} \cline{4-4}
\hline
                 & away from horizon & \checkmark & \checkmark & \checkmark & \checkmark \\
\hline
non-extremal & near-horizon  & \XSolidBrush & \XSolidBrush & \XSolidBrush & \XSolidBrush\\
\cline{2-2} \cline{4-4}
\hline
                 & away from horizon & \checkmark & \checkmark & \checkmark & \checkmark \\
\hline

\end{tabularx}
\caption{Table for pulsating string. The symbol \checkmark (\XSolidBrush) represents the non-violation (violation) of the bound.}\label{tab:str}
\end{table}

In future, we would like to address the following points


(1) The full analytical reasoning behind the violation of chaos bound for pulsating string is unknown to us and therefore in future it would be interesting to analyse this in more detail.

(2) The numerical evidence proposed in this work induces further analysis of the contrasting behaviour on the equivalence of the particle-string system proposed in \cite{PhysRevD.104.066017}.  

(3) 
It has been argued that the celebrated MSS bound can be extended even for asymptotically flat-spacetime \cite{ma_chaotic_2022}. Our work suggests something similar to this direction. In future, it would be interesting to explore this speculation with more rigour.

\section*{Appendix} \label{appendix}
\subsection*{A. Details of the numerics}
For all the calculations presented in the paper, we set $G_{10} = 1$, M = $\frac{5 \pi^{2}}{16}$, $\frac{2 \pi}{3}$, $\frac{3 \pi}{8}$, and $\frac{1}{2}$  for p = 3,4,5 and 6 respectively. With this, the charge and mass corresponding to black 3-brane satisfy the inequality $Q \leq \frac{8}{\pi^2}M$. The Hawking temperature is given by the relation
\begin{equation*}
    \beta =  \frac{1}{T} = \pi r_{+}[1-(\frac{r_{-}}{r_{+}})^4]^{-\frac{1}{4}}
\end{equation*}
Similarly, for p = 4, we find $Q \leq \frac{3}{\pi}M$ and the associated Hawking temperature is
\begin{equation*}
\beta =  \frac{1}{T} = \frac{4 \pi r_{+}}{3} [1-(\frac{r_{-}}{r_{+}})^3]^{-\frac{1}{6}}   
\end{equation*}
For the p = 5 and p = 6, the corresponding inequalities are 
\begin{equation*}
    Q \le 4 M/\pi, \hspace{3mm} Q \le 2 M 
\end{equation*}
respectively. Note that at the extremal limit, T = 0 for both p = 3 and p = 4, T = finite for p = 5 and $T\rightarrow \infty$ for p = 6.

Solving a nonlinear system of differential equations requires reliable numerical methods to control the error propagation. In this paper, we use the \texttt{Projection} method of \texttt{NDSolve} routine of \texttt{Mathematica} to solve the equations of motion. The constraint $|H|$ $< \delta$ (error tolerance) has been checked at every integration step with $\delta \sim 10 ^{-6}$. We perform the calculation of Lyapunov exponents using the variational algorithm described in \cite{sandri1996numerical}. \\
\newline
\subsection*{B. Fixed point} \label{fixed pt}
To analyze the large-time behaviour of a trajectory, it is helpful to study the possible invariant sets of the phase space of the dynamical system. The invariant set possesses the characteristic that any trajectory in phase space originating from a point within it will remain confined to it indefinitely. In this section, we obtain the fixed point corresponding to the system of equations \ref{eqn: energy2}-\ref{eqn: phieqn}. To do so, we set these equations equal to zero $(\dot r=0, \dot p_{r}=0, \dot \phi=0, \dot p_{\phi_{1}}=0)$, which provides the solution: $(r_{\ast}= r_{sol}, p_{\ast}= 0, \phi_{\ast}= (2N - 1)\frac{\pi}{2}, 0)$ where $r_{sol}$ is the solution of the equation :
\begin{align}
    (\gamma +1) \left(-n^2\right) r^2 \Delta _-^{\gamma } \Delta _+'(r)-\frac{n^2 r^2 \Delta _-^{\gamma +1} \Delta _-'(r)}{\Delta _+}-2 n^2 r \Delta _-^{\gamma +1}+\frac{E^2 \Delta _+'(r)}{2 \sqrt{\Delta _-} \Delta _+}=0
\end{align}
This also satisfies the Hamiltonian constraint $H=0$. For a specific choice of parameters, for instance: $p=3,~~ Q= 2.0, ~~ n=1, ~~E= 5.2$, we get $r^{\ast}=1.66435$. The corresponding Jacobian matrix at this fixed point has the eigenvalue $\{-3.37057, 3.37057, 1, -1 \} $, which shows that there is no attractor in our system as all the eigenvalues do not have negative real parts. Similarly, there is no repellor in the system as all the eigenvalues are not positive. 
Moreover, on increasing $n$, the threshold energy value for a specific set of parameters changes and generally increases. Similar behaviour has been observed for all $p<7$.
\newline
However, if we consider the large r limit in the system of equations  \ref{eqn: energy2}-\ref{eqn: phieqn}, there exists another fixed point ($r^{*} = \infty$, $p_{r}^{*} = 0$, $\phi_{1}^{*} = N\pi$, $p_{\phi_{1}}^{*}$ = arbitrary). The invariant set of the trajectories which are escaping to infinity has a measure of zero.\\
\newline
\subsection*{C. String trajectory}
\begin{figure}[h!]
\begin{subfigure}
  \centering
  \includegraphics[width=0.45\textwidth]{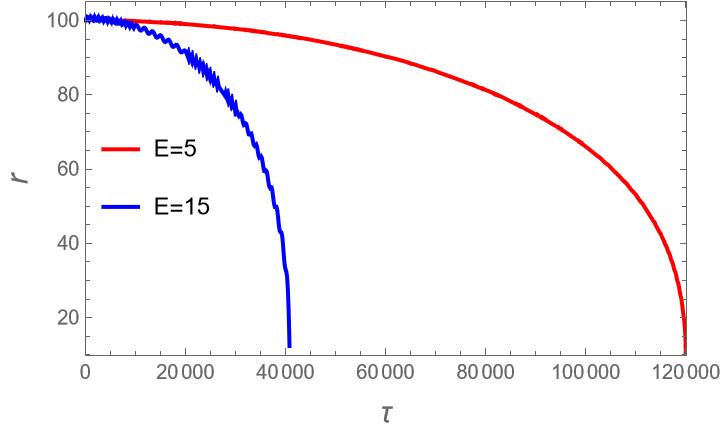}
\put(-182,123){(a)}
\end{subfigure}
\hspace{6mm}
\begin{subfigure}
  \centering
  \includegraphics[width=0.45\textwidth]{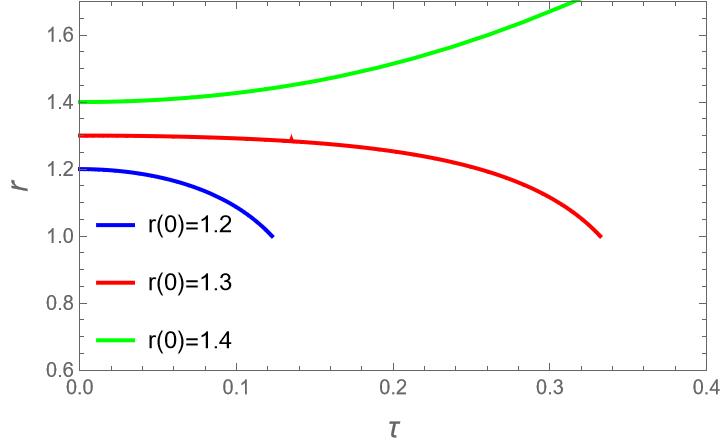}
\put(-182,123){(b)}
\end{subfigure}
\caption{String trajectory near p =3 brane. (a) $r(0)=100$. (b) $r(0) = 1.2 r_{h}, 1.3 r_{h}, 1.4 r_{h}$. For both the plots, we set E = 5, $P_{r}(0) = 0$, $\phi_{1} = 0$ and $r_{h} = 1$.}
\label{fig: trajectory(p=3)}
\end{figure}

\begin{figure}[h!]
\begin{subfigure}
  \centering
  \includegraphics[width=0.45\textwidth]{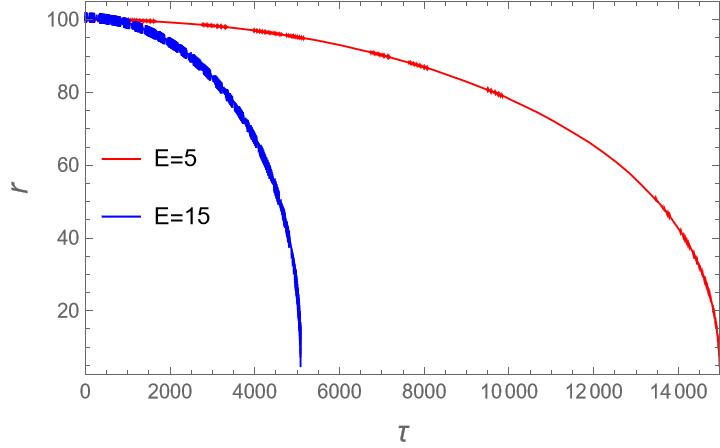}
\put(-182,123){(a)}
\end{subfigure}
\hspace{6mm}
\begin{subfigure}
  \centering
  \includegraphics[width=0.45\textwidth]{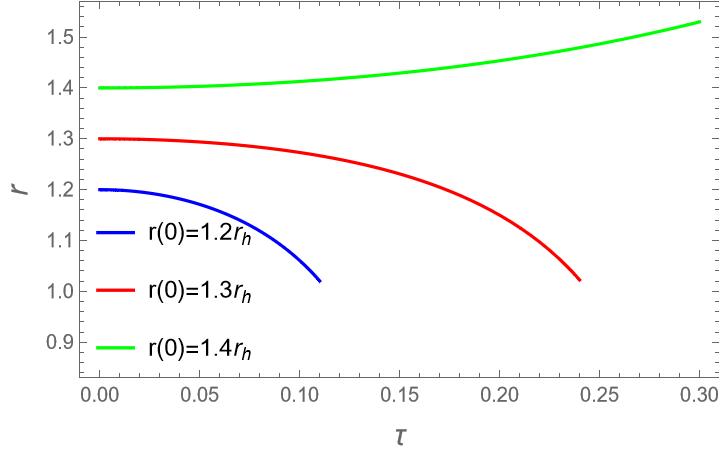}
\put(-182,123){(b)}
\end{subfigure}
\caption{String trajectory near p =4 brane. (a) $r(0)=100$. (b) $r(0) = 1.2 r_{h}, 1.3 r_{h}, 1.4 r_{h}$. For both the plots, we set E = 5, $P_{r}(0) = 0$ and $\phi_{1} = 0$ and $r_{h} = 1$.}
\label{fig: trajectory(p=4)}
\end{figure}

In this section, we provide the string trajectory (p=3 and p=4) by numerically solving the coupled equations of motion \ref{eqn: energy2}-\ref{eqn: phieqn}. 
%
Without loss of generality, we assume Q = 0. First, we consider the  neutral 3-brane with different initial locations of the string - far from the horizon, 
 and near the horizon. 
 We consider the scenario when the string initially is far from the horizon ($r(0) = 100$). For $p_{r}(0) = 0$, we observe the capture of the string with any value of E (fig \ref{fig: trajectory(p=3)}(a)). This is due to the dominance of the attractive potential (-$\frac{E^2}{2 \Delta_{+}\Delta_{-}^{-1/2}}$) term. At sufficiently high E, the string is always likely to get captured. Note that the capture time decreases with E. The string dynamics becomes even more complex when it is placed initially close to the horizon. When r(0)=$1.2r_{h}$, $1.3r_{h}$ and $p_{r}(0) = 0$, it gets quickly captured in the brane (fig \ref{fig: trajectory(p=3)}(b)). However, the string escapes to infinity for $r(0) = 1.4 r_{h}$.
 Next, we repeat the same exercise in p = 4 brane. Our numerical analysis reveals similar features in the string dynamics irrespective of the initial location of the string (fig \ref{fig: trajectory(p=4)}). However, the capture of the strings occurs quicker than the corresponding p = 3 brane. 

\bibliographystyle{JHEP}
\bibliography{chaosbound}
\end{document}